\documentclass[12pt,letterpaper]{article}
\usepackage{jheppub}
\usepackage{verbatim}
\usepackage{slashed}

\usepackage{graphicx}
\usepackage{subfigure}
\input{epsf}
\usepackage{epsfig}
\usepackage{epstopdf}
\usepackage{amsthm}

\usepackage{tikz}
\def\({\left(} \def\){\right)}
\def\[{\left[} \def\]{\right]}

\newcommand{\be}{\begin{equation}}
\newcommand{\ee}{\end{equation}}
\newcommand{\bea}{\begin{eqnarray}}
\newcommand{\eea}{\end{eqnarray}}
\newcommand{\ba}{\begin{eqnarray}}
\newcommand{\ea}{\end{eqnarray}}

\newcommand{\beq}{\begin{equation}}
\newcommand{\eeq}{\end{equation}}
\newcommand{\beqa}{\begin{eqnarray}}
\newcommand{\eeqa}{\end{eqnarray}}
\newcommand{\beqar}{\begin{eqnarray*}}
\newcommand{\eeqar}{\end{eqnarray*}}

\newcommand{\eg}{{\it e.g.}\ }
\newcommand{\ie}{{\it i.e.}\ }
\title{Comments on Single-Trace $T\bar T$ Holography}

\author[a]{Soumangsu Chakraborty}
\author[b]{, Amit Giveon}
\author[c]{and  David Kutasov}
\affiliation[a]{Institut de Physique Th\'eorique, Universit\'e Paris-Saclay, CNRS, CEA\\	
Orme des Merisiers, 91191 Gif-sur-Yvette, France}
\affiliation[b]{Racah Institute of Physics, The Hebrew University, Jerusalem, 91904, Israel}
\affiliation[c]{Kadanoff Center for Theoretical Physics, Enrico Fermi Institute, and Department of Physics  \\University of Chicago, 5640 S. Ellis Ave, Chicago, IL 60637, USA}

\emailAdd{soumangsuchakraborty@gmail.com}
\emailAdd{giveon@mail.huji.ac.il}
\emailAdd{dkutasov@uchicago.edu}

\abstract{We explore the holographic duality between string theory in backgrounds that interpolate between asymptotically linear dilaton spacetime in the UV and $AdS_3$ in the IR, and single-trace $T\bar T$ deformed CFT. In particular, we explain how the deformation of states in the boundary theory is reflected in the bulk geometry, and show that the coupling above which the deformed energy of the $SL(2,\mathbb{R})$ invariant ground state of the IR CFT becomes complex is a maximal coupling.}

\begin{document}
\maketitle
 
 \section{Introduction}\label{sec1}

$T\bar T$ deformed conformal field theories (see \eg \cite{Jiang:2019epa} for a review), and their various generalizations (see \eg \cite{Chakraborty:2019mdf} and references therein), are of interest for a number of reasons. One is that they correspond to irrelevant deformations of (two-dimensional) CFT's, in which, in some cases, one can flow up the RG without encountering ambiguities and/or singularities, unlike the general behavior of QFT in the presence of such deformations. 

Moreover, these theories do not approach a fixed point of the RG in the UV. Rather, in a particular range of parameters, they exhibit Hagedorn behavior of the entropy at high energy, which is characteristic of non-local theories like string theory. Thus, they provide a playground for studying non-local theories using the tools of QFT. 

Another reason for the interest in these theories has to do with holography. If the original CFT has an $AdS_3$ dual, it is natural to ask what is the effect of the $T\bar T$ deformation on it. Under the standard AdS/CFT dictionary, $T\bar T$ is a ``double-trace'' operator, since it is a product of two stress-tensors, each of which has a bulk dual (see \eg \cite{Brown:1986nw,Kutasov:1999xu,Aharony:1999ti}). There has been extensive discussion of this double-trace deformation, starting with \cite{McGough:2016lol}. In these models, the bulk geometry remains $AdS_3$, and the deformation is encoded in the boundary conditions of the bulk fields, see \eg \cite{Guica:2019nzm}. 

A second class of holographic models related to $T\bar T$ deformed CFT is the single-trace deformation introduced in \cite{Giveon:2017nie}. On the boundary, a rough way of thinking about these models is the following. If the undeformed CFT is a symmetric product, ${\cal M}^p/S_p$, the double-trace $T\bar T$ deformation mentioned above takes the form $\delta L=tT\bar T$, where $T=\sum_i T_i$ is the total stress tensor of the theory. The single-trace deformation is $\delta L=t\sum_i T_i\bar T_i$, the sum of $T\bar T$ deformations in the blocks. 

The spacetime CFT corresponding to the most well-studied $AdS_3$ backgrounds, \eg $AdS_3\times S^3\times T^4$, is not a symmetric product, but the two have much in common \cite{Argurio:2000tb,Giveon:2005mi,Chakraborty:2019mdf}, so the symmetric product picture provides a useful guide to the physics of single-trace $T\bar T$ deformed CFT. Below the $k=1$\footnote{$k$ is related to the radius of curvature of $AdS_3$, $R_{\rm AdS}=\sqrt{k} l_s$.} string-black hole transition of \cite{Giveon:2005mi}, the symmetric product picture becomes much more accurate \cite{Balthazar:2021xeh,Eberhardt:2021vsx}. 

Unlike its double-trace cousin, single-trace $T\bar T$ deformed holographic CFT is described by a bulk background that differs from $AdS_3$ \cite{Giveon:2017nie}. The deviation from $AdS_3$ grows as one increases the radial coordinate, in agreement with the fact that the deformation is irrelevant. In the cases where the asymptotic density of states is Hagedorn, the background approaches a linear dilaton spacetime near the boundary \cite{Giveon:2017nie,Chakraborty:2019mdf}. 

As is standard in holographic dualities, one can use the boundary theory to learn about the bulk one and vice versa. In single-trace $T\bar T$ deformed CFT, one way to do the former is to {\it define} the bulk string theory in a class of asymptotically linear dilaton spacetimes via the dual deformed CFT's, thereby extending the range of models where quantum gravity can be understood non-perturbatively using holography. This is particularly useful for the $k<1$ models of \cite{Balthazar:2021xeh,Eberhardt:2021vsx}, but there are reasons to believe that the construction can be extended to $k>1$ backgrounds, like $AdS_3\times S^3\times T^4$. 

An example where one can use the bulk theory to shed light on the boundary one was discussed in \cite{Chakraborty:2020swe,Chakraborty:2020cgo}. As is well known \cite{Smirnov:2016lqw,Cavaglia:2016oda}, if the original CFT is placed on a circle $S^1_x$ of radius $R$, a state in the CFT of energy $E$ and momentum $P$ gives rise in the $T\bar T$ deformed theory to a state of energy $ E(\lambda)$ (and the same momentum), with 
\begin{equation}\label{TTE}
 E(\lambda)=\frac{1}{\lambda R}\left[\sqrt{1+2\lambda RE(0)+(\lambda RP)^2}-1\right],
 \end{equation}
where $E(0)\equiv E$. $\lambda$ is related to the dimensionful $T\bar T$ coupling $t$, $\lambda\sim t/R^2$. It can be thought of as the value of the coupling $t$ at the ``Kaluza-Klein'' scale $R$. 

If one fixes the dimensionful coupling $t$, and takes the radius of the spatial circle, $R$ to infinity, the coupling $\lambda$ goes to zero, and the deformed energies (more precisely $ER$, which are related to the scaling dimensions of the corresponding operators in the IR CFT) approach their undeformed values. For large but finite $R$, $\lambda\ll 1$, and eq. \eqref{TTE} behaves as follows. For\footnote{We will set the momentum on $S^1_x$ to zero here and below, for simplicity. The generalization  to non-zero momentum appears in \cite{Chakraborty:2023zdd}.} $\lambda R E\ll1$, the deformed energies $E(\lambda)$ are approximately equal to the undeformed ones, $E(\lambda)\simeq E(0)$. The deformation becomes significant for states with energies around $E_{\rm c}$, defined as
\begin{equation}\label{crossover}
 \lambda R E_{\rm c}= 1~.
 \end{equation}
One can think about the energy $E_{\rm c}$ as follows (see \cite{Giveon:2017nie} for further discussion).  

Since $RE_c= 1/\lambda\gg1$, the undeformed CFT has many states with $1\ll RE\ll RE_c$. For such energies, the CFT entropy can be approximated by the Cardy formula $S_{\rm CFT}\sim\sqrt{ER}$, up to an overall constant that depends on the central charge of the CFT, $c$. On the other hand, for energies much above $E_{\rm c}$ the spectrum of states is Hagedorn, with entropy $S_{\rm H}\sim \sqrt t E$, again, up to a $c$-dependent constant. One can think of \eqref{crossover} as the energy at which the crossover between the two behaviors takes place; \ie\ it can be obtained by setting $S_{\rm CFT}=S_{\rm H}$. For $E\ll E_{\rm c}$ the entropy of the deformed theory is given by the CFT entropy, while for $E\gg E_{\rm c}$ it is given by the Hagedorn one.\footnote{This is reminiscent of the discussion of the string/black hole transition \cite{Horowitz:1996nw}, where the roles of the CFT and Hagedorn phases are played by fundamental strings and large black holes, respectively.} 

Equation \eqref{TTE} also implies that the energy $E(\lambda)$ can become complex for negative coupling, $\lambda<0$, and sufficiently large undeformed energy $E$. This happens when $RE>1/(2|\lambda|)$. There has been much discussion in the literature on whether this means that theories with negative $\lambda$ are inconsistent;\footnote{Interestingly, the  holographic construction of \cite{McGough:2016lol} was for this sign of the coupling.} see \eg\ \cite{Aharony:2018bad}. In \cite{Chakraborty:2020swe,Chakraborty:2020cgo} we studied this question in single-trace $T\bar T$ deformed CFT, using the holographic construction of \cite{Giveon:2017nie}. The bulk spacetime is singular in this case, but analyzing the behavior of black holes and fundamental strings in this background we did not find any obvious signs of inconsistency. 

Another interesting feature of \eqref{TTE} is that if we take the coupling $\lambda$ to be positive, the energy $ E(\lambda)$ becomes complex if the undeformed energy is small enough, and the coupling $\lambda$ large enough. In particular, for the $SL(2,\mathbb{R})$ invariant ground state of the undeformed CFT, which has $ER=-c/12$, the energy $ E(\lambda)$ becomes complex for $c\lambda>6$. In this note, we will discuss the question of what happens in single-trace $T\bar T$ deformed CFT when $\lambda$ approaches this value. This question was raised in \cite{Apolo:2019zai,Chang:2023kkq,Benjamin:2023nts}. We will show that from the bulk point of view, the critical value of $\lambda$ serves as a maximal coupling, and discuss the interpretation of this fact in the boundary theory. 

We will also use the opportunity to make some additional comments on the holographic duality between single-trace $T\bar T$ deformed CFT and string theory in asymptotically linear dilaton spacetime. In particular, we will discuss how the deformation of states in the boundary theory is reflected in the corresponding bulk geometry, and derive the appropriate version of \eqref{TTE} for the single-trace case directly from the bulk perspective.

 \section{The background with $M\geq 0$}\label{sec2}

The starting point of our discussion is type II string theory on $\mathbb{R}_t\times S^1_x\times T^4\times\mathbb{R}^4$, with $k$-NS5 branes wrapping $S^1_x\times T^4$ and $p$ fundamental strings wrapping $S^1_x$. We denote the radius of the $x$ circle by $R$, and the volume of the $T^4$ by $V=(2\pi l_s)^4v$. We will be interested in the theory living on the intersection of the strings and the fivebranes, \ie\ on the $1+1$ dimensional spacetime labeled by $(t,x)$. 

In the ground state, the above brane configuration preserves eight supercharges. Its energy is given by   
\begin{equation}\label{extE}
 E_{\rm ext}=p\frac{R}{l_s^2}+k\frac{Rv}{g^2ls^2}~.
 \end{equation}
We will consider non-extremal configurations, which are described by the black brane background \cite{Maldacena:1996ky,Hyun:1997jv}
 \begin{equation}
 \begin{aligned}\label{ns5f1n}
 &ds^2=\frac{1}{f_1}\left(-fdt^2+dx^2\right)+f_5\left(\frac{1}{f}dr^2+d\Omega_3^2\right)+\sum_{i=1}^4 dx_i^4~,\\
 &e^{2\Phi}=g^2\frac{f_5}{f_1}~,\\
 &H=dx\wedge dt \wedge d\left(\frac{r_0^2\sinh2\alpha_1}{2f_1r^2}\right)+r_0^2\sinh2\alpha_5d\Omega_3~,
 \end{aligned}
 \end{equation}
 where $(x_1,\cdots, x_4)$ are coordinates on $T^4$, $(r,\Omega_3)$ are spherical coordinates on the transverse $\mathbb{R}^4$, $r_0$ is the radial position of the horizon, and $g=e^{\Phi(r\to\infty)}$ is the string coupling far from the fivebranes. It is related to the ten-dimensional Newton constant via
 \begin{equation}\label{GNten}
 G_N^{(10)}=8\pi^6g^2l_s^8~.
 \end{equation}
 
 The various harmonic functions in \eqref{ns5f1n} are given by 
 \begin{equation}\label{harfun}
 f=1-\frac{r_0^2}{r^2}~,\ \ \ \  f_{1,5}=1+\frac{r_{1,5}^2}{r^2}~, \ \ \ \ r_{1,5}^2=r_0^2\sinh^2\alpha_{1,5}~,
 \end{equation}
 with
 \begin{equation}\label{charges}
 \sinh2\alpha_1= \frac{2p g^2l_s^2}{vr_0^2}~, \ \ \ \   \sinh2\alpha_5=\frac{2l_s^2k }{r_0^2}~.
 \end{equation}
 Equations \eqref{harfun}, \eqref{charges} imply that $r_1$, the crossover scale between the two asymptotic behaviors of the harmonic function $f_1$, depends on $r_0$, the radial position of the horizon of the black hole (a fact that played an important role in \cite{Chakraborty:2020swe,Chakraborty:2020cgo}). 
The extremal configuration is obtained by taking $r_0\to 0$. As is clear from \eqref{charges}, in this limit $\alpha_1\to\infty$, with $r_0\exp(\alpha_1)$ held fixed. Thus, $r_1$ approaches the value 
\begin{equation}\label{rone}
     r_1^2= pg^2l_s^2/v~.
 \end{equation}
The background \eqref{ns5f1n} is asymptotically flat. Its ADM mass is \cite{Maldacena:1996ky}
 \begin{equation}\label{madm1}
 M_{ADM}=\frac{Rvr_0^2}{2l_s^4g^2}(\cosh2\alpha_1+\cosh2\alpha_5+1)~.
 \end{equation}
The energy above extremality is given  by
 \begin{equation}\label{eaext}
 E=M_{ADM}-E_{\rm ext}=\frac{Rvr_0^2}{2l_s^4g^2}(e^{-2\alpha_1}+e^{-2\alpha_5}+1)~,
 \end{equation}
where we used \eqref{extE}, \eqref{charges}, \eqref{madm1}.

The decoupled fivebrane theory is obtained by taking the limit $g\to 0$, with $r$, $r_0$, $r_1$ scaling like $g$ \cite{Aharony:1998ub}. To achieve that, we rescale 
 \begin{equation}\label{declim}
     r\to g r~, \ \ \ r_0\to g r_0~,  \ \ \ r_1\to g r_1
 \end{equation}
 in \eqref{ns5f1n}, and take 
 \begin{equation}\label{gto0}
     g\to 0
 \end{equation}
while holding the rescaled $r$, $r_0$, $r_1$ fixed.
Equations \eqref{harfun}, \eqref{charges} imply that in this limit $\alpha_5\to \infty$, and $r_5$ approaches the value
\begin{equation}\label{r5}
     r_5=\sqrt{k}l_s~.
 \end{equation}
The energy above extremality \eqref{eaext} takes the form 
\begin{equation}\label{eextm3}
    E=\frac{Rvr_0^2}{2l_s^4}\left(e^{-2\alpha_1}+1\right).
\end{equation}
The background \eqref{ns5f1n} factorizes in the limit \eqref{declim}, \eqref{gto0} into the product of a $T^4$, an $S^3$ of radius $r_5$, \eqref{r5}, described by a level $k$ supersymmetric WZW model,  and a non-trivial $2+1$ dimensional spacetime corresponding to the background fields 
\begin{equation}\label{ns5f1dec}
  \begin{aligned} 
  &ds^2=\frac{1}{f_1}\left(-fdt^2+dx^2\right)+\frac{r_5^2}{r^2f}dr^2~,\\
 &e^{2\Phi}=\frac{r_5^2}{r^2f_1}~,\\
 &H=dx\wedge dt \wedge d\left(\frac{r_0^2\sinh2\alpha_1}{2r_5^2}e^{2\Phi}\right)~,
 \end{aligned}
 \end{equation}
 where $f,f_{1}$ are given by \eqref{harfun}, with \eqref{charges}, \eqref{declim}, 
 \begin{equation}\label{alpha1}
     \sinh 2\alpha_1=\frac{2l_s^2p}{vr_0^2}~.
 \end{equation}
 In the extremal case ($r_0=0$), the background \eqref{ns5f1dec} interpolates between massless BTZ with $R_{\rm AdS}=r_5$, \eqref{r5}, for $r\ll r_1=l_s\sqrt{p/v}$, \eqref{rone}, \eqref{declim}, and $2+1$ dimensional flat spacetime with linear dilaton in the radial direction for $r\gg r_1$. This background is referred to in \cite{Giveon:1999zm,Giveon:2017nie} as $\mathcal{M}_3$. String theory in it is holographically dual to a boundary theory that interpolates between a two-dimensional CFT in the IR and a non-local theory with a Hagedorn density of states, known as Little String Theory (LST), \cite{Aharony:1998ub}, in the UV. As described in \cite{Giveon:2017nie}, it can also be thought of as an irrelevant deformation of the IR CFT by a single-trace analog of the $T\bar T$ operator. 

For $r_0>0$, \eqref{ns5f1dec} describes {\it states} in the above single-trace $T\bar T$ deformed CFT. The energies of these states are related to the radial position of the horizon by \eqref{eextm3}, \eqref{alpha1}. In order to study such states, it is useful to change coordinates from those used in \eqref{ns5f1dec}. The radial coordinate $r$, in particular, is inconvenient since, as discussed in \cite{Chakraborty:2020swe}, in terms of this coordinate, the parameter $r_1$ in eq. \eqref{harfun} depends on the location of the black hole horizon, $r_0$. Since $r=r_1$ is the radial position where the geometry makes a transition from $AdS_3$ to linear dilaton, this makes the analysis cumbersome. 

To circumvent this problem (and some others), we change coordinates from $(t,r,x)$ to $(\tau,\rho,\varphi)$,
 \begin{equation}\label{coorbtz}
     \begin{aligned}
         \tau=\frac{r_5}{R}t ~,\ \ \ \ 
         \rho=\frac{Rr}{r_0\sinh\alpha_1}~, \ \ \ \
         \varphi=\frac{x}{R}~.
     \end{aligned}
 \end{equation}
The rescaled spatial boundary coordinate is identified as  $\varphi\sim\varphi+2\pi$, and $r_5$ is given by \eqref{r5}.

In the new coordinates, the background \eqref{ns5f1dec} takes the form
\begin{equation}\label{fullbgrho}
    \begin{aligned} 
        &ds^2=-\frac{N^2}{1+\frac{\rho^2}{R^2}}d\tau^2+\frac{d\rho^2}{N^2}+\frac{\rho^2}{1+\frac{\rho^2}{R^2}}d\varphi^2~,\\
      &  B_{\tau\varphi}=\frac{\rho^2}{r_5}\sqrt{1+\frac{\rho_0^2}{R^2}}\frac{1}{1+\frac{\rho^2}{R^2}}~,\\
      & e^{2\Phi}= \frac{kv}{p}\sqrt{1+\frac{\rho_0^2}{R^2}}\frac{1}{1+\frac{\rho^2}{R^2}}~,
    \end{aligned}
\end{equation}
where
\begin{equation}\label{Nfull}
    N^2=\frac{\rho^2-\rho_0^2}{r_5^2}
\end{equation}
and, \eqref{coorbtz},
\begin{equation}\label{rho0full}
    \rho_0=\frac{R}{\sinh\alpha_1}~.
\end{equation}
The geometry \eqref{fullbgrho} -- \eqref{rho0full} describes a black hole with a horizon at $\rho=\rho_0$. Its Bekenstein-Hawking entropy and energy, \eqref{eextm3}, are given by
\begin{equation}\label{entropy}
    S=2\pi kp\frac{\rho_0}{r_5}
\end{equation}
and
\begin{equation}\label{EE}
    E=\frac{Rp}{l_s^2}\left(-1+\sqrt{1+\frac{\rho_0^2}{R^2}}\right),
\end{equation}
where we used the relation 
\begin{equation}\label{r0rgo0}
    r_0^2=\frac{l_s^2p \rho_0^2}{R^2v\sqrt{1+\frac{\rho_0^2}{R^2}}}
\end{equation}
that follows from \eqref{alpha1}, \eqref{rho0full}. 

The background \eqref{fullbgrho} -- \eqref{rho0full} has three scales: the radius of $S^1_x$, $R$, the location of the horizon, $\rho_0$, and the radius of curvature of the infrared $AdS_3$, $r_5$, \eqref{r5}. 
From the point of view of the boundary theory, one can think of them as the size of the spatial circle, the energy of the state \eqref{EE}, and the single-trace $T\bar T$ coupling, $t\sim l_s^2$, \cite{Giveon:2017nie}. 

This identification allows one to write the energy \eqref{EE} in a way familiar from studies of $T\bar T$ deformed CFT. To do that, it is convenient to define the dimensionless single-trace $T\bar T$ coupling \cite{Giveon:2017nie}
\begin{equation}\label{lambda}
    \lambda\equiv \frac{l_s^2}{R^2}\;. 
\end{equation}
As mentioned in section \ref{sec1}, $\lambda$ can be thought of as the value of the single-trace $T\bar T$ coupling at the scale $R$.   

As discussed above, one can think of \eqref{fullbgrho} -- \eqref{rho0full} as describing states in a theory labeled by the coupling $\lambda$. For $\lambda= 0$, these are states in the  undeformed CFT, which is described in the bulk by an asymptotically $AdS_3$ background. As we change $\lambda$, the states are deformed, and in particular, their energy \eqref{EE} becomes a function of $\lambda$, $E=E(\lambda)$.  

In order to compute $E(\lambda)$, we need to understand how $\rho_0$ in \eqref{fullbgrho} changes with $\lambda$ for a given state in the undeformed theory. Assuming that the deformed energy is only a function of the undeformed energy and the coupling, as in \cite{Aharony:2018bad}, the entropy must satisfy
\begin{equation}\label{deformentropy}
    S(\lambda, E(\lambda))=S(0, E(0))~.
\end{equation}
Looking back at \eqref{entropy}, we conclude that if we want to follow the same state as we vary $\lambda$, we need to hold the horizon radius $\rho_0$ fixed, 
\begin{equation}\label{rholambda}
\rho_0(\lambda)=\rho_0(0)\equiv\rho_0~.
\end{equation}
To determine $\rho_0$, we take the limit to the undeformed theory, $\lambda\to 0$. In the bulk description, this corresponds to taking $R\to\infty$, \ie\ considering the background \eqref{fullbgrho} -- \eqref{rho0full} in the limit $\rho_0,\rho\ll R$. In this limit, the background takes the form  
  \begin{equation}\label{btxmetric}
 \begin{aligned}
   &  ds^2=-N^2d\tau^2+\frac{d\rho^2}{N^2}+\rho^2d\varphi^2~,\\
 &  B_{\tau\varphi}=\frac{\rho^2}{r_5}~,\\
 &  e^{2\Phi}=\frac{kv}{p}~,  
 \end{aligned}
 \end{equation}
 with
 \begin{equation}\label{NNphi}
     N^2=\frac{\rho^2-\rho_0^2}{r_5^2}\equiv\frac{\rho^2}{r_5^2}-8G_3M~,
 \end{equation}
 which describes a BTZ black hole of mass $M$, in an $AdS_3$ spacetime with $R_{\rm AdS}=r_5$, \eqref{r5}, horizon position $\rho_0$, and three-dimensional Newton constant, $G_3$, given by
\begin{equation}\label{G3rho0}
   \rho_0^2=8 r_5^2 G_3 M~, \ \ \ G_3=\frac{l_s}{4\sqrt{k}p}~.
\end{equation}
This black hole corresponds in the boundary CFT to a state with energy (see \eg\ \cite{Kraus:2006wn} for a review)
\begin{equation}\label{E0}
     E=\frac{r_5}{R} M~.
 \end{equation}
From the point of view of the preceding discussion, the energy \eqref{E0} can be thought of as $E(\lambda=0)$. 
The deformed energy $E(\lambda)$ is obtained by plugging \eqref{lambda}, \eqref{G3rho0}, \eqref{E0}  into \eqref{EE}, which gives
\begin{equation}\label{zamo}
     \frac{1}{p}E(\lambda)=\frac{1}{\lambda R}\left(-1+\sqrt{1+2\lambda RE(0)/p}\right).
 \end{equation}

An interesting feature of the above discussion is that equation \eqref{zamo} has the same qualitative form as \eqref{TTE}. In fact, the same equation would be obtained if we studied a symmetric product CFT, and: (a) $T\bar T$ deformed the building block of the symmetric product; (b) considered states in which the energy is divided equally between all $p$ factors. 

This raises the questions: (a) why does the energy formula look like it comes from a symmetric product, when the boundary CFT is not one, as mentioned above; (b) why is the energy divided equally among the $p$ factors? The answer to these questions requires a better understanding of the boundary CFT dual to string theory on $AdS_3$ (with $k>1$\footnote{As mentioned above,  for $k<1$ the boundary CFT is much better understood, \cite{Balthazar:2021xeh,Eberhardt:2021vsx}, but the issue discussed in this section does not exist, since that theory does not contain black holes, \cite{Giveon:2005mi}.}) than is currently available. The  status of the symmetric product appears to be the following. 

The situation is clearest for states with energies that go like $p^0$ in the large $p$ limit. These are perturbative string states, and they belong to one of two classes: normalizable states, that belong to the principal discrete series, and delta-function normalizable states, that belong to the principal continuous series. In both cases, these states can carry winding \cite{Maldacena:2000hw}.  

The spectrum of principal continuous series states, often referred to as {\it long strings}, is known to be well described by a symmetric product of the form $\left({\cal M}_{6k}\right)^p/S_p$, where ${\cal M}_{6k}$ is a CFT with central charge $c_{\cal M}=6k$ corresponding to long string states with winding $w=1$ (see \eg \cite{Chakraborty:2019mdf} and references therein). That CFT has a Liouville structure, \cite{Seiberg:1999xz}, and in particular its effective central charge $c_{\rm eff}$ \cite{Kutasov:1990sv,Kutasov:1991pv} is smaller than $c_{\cal M}$, $c_{\rm eff}=6\left(2-\frac1k\right)$, \cite{Giveon:2005mi}. More precisely, the agreement is between states with energies that scale like $p^0$ in the above symmetric product CFT and long string states with winding $w$ that scale like $p^0$. 

Principal discrete series states are known {\it not} to fit into the above symmetric product. However, these states have measure zero in the string spectrum, so one can say that the symmetric product describes generic states in perturbative string theory on $AdS_3$. As the energy of long strings increases, their coupling does as well, as discussed in \cite{Giveon:2005mi}, following \cite{Seiberg:1999xz}. Eventually, for energies of order $p$, they give rise to BTZ black holes. 

For $k>1$, it appears that the symmetric product structure cannot persist for arbitrary energy. Indeed, suppose it did, perhaps with the seed ${\cal M}_{6k}$ replaced by a different one ${\cal \widetilde M}_{6k}$. Since it is known that\footnote{For $k>1$; see \cite{Giveon:2005mi}.} the $SL(2,\mathbb{R})$ invariant ground state of the boundary CFT must be in the spectrum, the CFT ${\cal \widetilde M}_{6k}$ must have this property as well. Modular invariance and unitarity then imply that this CFT has a much richer spectrum than that on ${\cal M}_{6k}$, which is inconsistent with the fact that the extra states have energies of order $p^0$, and those should be seen in the spectrum of perturbative string states, which are captured by the CFT ${\cal M}_{6k}$.\footnote{A possibility, discussed in \cite{Eberhardt:2021vsx}, is that the seed of the symmetric product only contains the principal continuous states, and the discrete states are due to the existence of a Liouville type potential that goes to zero at a large radial distance, and involves twisted sector operators, as in the $k<1$ construction of \cite{Balthazar:2021xeh}. It is currently not understood whether this proposal is consistent. One difficulty with it is that to be compatible with \cite{Seiberg:1999xz}, the strong coupling region cannot be shielded by the potential. Another is modular invariance at finite $p$.}

Equation \eqref{zamo} seems to suggest that at very high energies, where the black hole description is valid, the symmetric product picture becomes reliable again, at least for generic states. It is not currently understood why that should be the case; it would be interesting to understand this better. 

Another important feature of the above discussion is the correspondence between the action of the deformation on the bulk geometry and on the boundary theory. In the bulk, the difference between the deformed energy \eqref{EE} and the undeformed one \eqref{E0} is due to the fact that as we decrease $R$, it approaches the horizon $\rho_0$, which means that the horizon of the black hole approaches the region where the crossover from $AdS_3$ to linear dilaton spacetime takes place. 

In the boundary theory, the difference between \eqref{zamo} and $E(0)$ is due to the fact that as the coupling $\lambda$ \eqref{lambda} increases, the undeformed energy approaches the energy at which the transition between Cardy and Hagedorn growth of the density of states takes place. This was discussed in section \ref{sec1}, around equation \eqref{crossover}. 

There is thus a nice correspondence between the bulk black hole geometries and the corresponding boundary states. For $\rho_0<R$, the part of the bulk geometry \eqref{fullbgrho} with $\rho_0<\rho<R$ looks approximately like the exterior of a BTZ black hole \eqref{btxmetric}, \eqref{NNphi}, with the approximation becoming better as $\rho_0/R$ decreases, and worse as it approaches one. In the boundary theory, states with energies $E(0)\ll pE_c$, where $E_c$ is defined in eq. \eqref{crossover}, look approximately like states in the unperturbed CFT, with the approximation becoming better as $E(0)$ decreases, and worse as it approaches $pE_c$, \eqref{zamo}.

Similarly, for states with $\rho_0>R$, the region $\rho>\rho_0$ in \eqref{fullbgrho} looks approximately like the geometry outside the two-dimensional black hole $SL(2,\mathbb{R})/U(1)$ in the $(\tau,\rho)$ directions, times $S^1_\varphi$, with the approximation becoming better as $\rho_0$ increases, and worse as it approaches $R$ (from above). In the boundary theory, the corresponding states have energies $E(\lambda)\simeq \sqrt{2pE(0)/\lambda R}$, with the approximation becoming better as $E(0)$ grows and worse as it approaches $pE_c$ from above.

\section{Deformation of global $AdS_3$}

We now come to one of the main points of this note. In section \ref{sec2} we discussed the bulk backgrounds corresponding to black holes in single-trace $T\bar T$ deformed CFT, \eqref{fullbgrho} -- \eqref{rho0full}. These formulae can be used to analyze the effects of the deformation on the $SL(2,\mathbb{R})$ invariant ground state of the boundary CFT, which is holographically dual to global $AdS_3$. To do that, we set the mass $M$ in eq. \eqref{fullbgrho}, \eqref{Nfull} to $8G_3M=-1$, or equivalently \eqref{G3rho0} $\rho_0^2=-r_5^2$. This leads to 
\begin{equation}\label{gads}
    \begin{aligned} 
        &ds^2=-\frac{N^2}{1+\frac{\rho^2}{R^2}}d\tau^2+\frac{d\rho^2}{N^2}+\frac{\rho^2}{1+\frac{\rho^2}{R^2}}d\varphi^2~,\\
      &  B_{\tau\varphi}=\frac{\rho^2}{r_5}\sqrt{1-\frac{r_5^2}{R^2}}\frac{1}{1+\frac{\rho^2}{R^2}}~,\\
      & e^{2\Phi}= \frac{kv}{p}\sqrt{1-\frac{r_5^2}{R^2}}\frac{1}{1+\frac{\rho^2}{R^2}}~,
    \end{aligned}
\end{equation}
where
\begin{equation}\label{Ngads}
    N^2=1+\frac{\rho^2}{r_5^2}~.
\end{equation}
Note that, unlike the discussion around eq. \eqref{rholambda}, here there is no ambiguity in determining $\rho_0(\lambda)$. From the bulk point of view, this is due to the fact that for negative $\rho_0^2$ in \eqref{fullbgrho}, its value is uniquely determined by the requirement that the geometry is smooth (no conical singularity) at $\rho=0$. 

From the boundary point of view, it is due to the fact that the undeformed energy of the $SL(2,\mathbb{R})$ invariant ground state is fixed to $RE(0)=-c/12$. In contrast, in section \ref{sec2} we discussed states that could have any (non-negative) energy in the undeformed theory, which led to the ambiguity noted around eq. \eqref{rholambda}. Thus, for the background \eqref{gads}, \eqref{Ngads}, equation \eqref{zamo} for the deformed energy does not rely on any assumptions about the dynamics of the deformed theory. It is obtained by plugging in $RE(0)=-kp/2(=-c/12)$ into \eqref{zamo}. 

In the limit $R\to\infty$, the background \eqref{gads}, \eqref{Ngads} approaches global $AdS_3$ for all $\rho$. For finite but large $R$ (more precisely, $R\gg r_5$), it interpolates between global $AdS_3$ for $\rho\ll R$ and a linear dilaton spacetime for $\rho\gg R$, with a smooth crossover at $\rho\sim R$. As $\rho/R$ increases, the size of the deformation grows, and the size of the region where \eqref{gads}, \eqref{Ngads} looks like global $AdS_3$ shrinks. As $R\to r_5$, the $B$-field and string coupling, $\exp(\Phi)$, go to zero for all $\rho$ and, as we will see below, the size of the $AdS_3$ region goes to zero. 

The value $R=r_5$ has a simple boundary interpretation. It corresponds to the $T\bar T$ coupling $\lambda=1/k$ (see \eqref{r5}, \eqref{lambda}), which is the value above which the energy of the $SL(2,\mathbb{R})$ invariant vacuum of the deformed boundary CFT becomes complex, see \eqref{zamo}. 

At first sight, the situation described above seems puzzling. The background \eqref{gads} corresponds in the boundary theory to an RG flow between a $CFT_2$ in the IR and LST in the UV. Thus, at low energies, it should approach the CFT. However, we find that the $SL(2,\mathbb{R})$ invariant vacuum {\it is} influenced by the deformation, even to the point that its energy can become complex at strong enough coupling. This seems surprising -- what can be more low energy than the vacuum!?

To understand the origin of this phenomenon, it is useful to consider a more conventional RG flow, that connects two fixed points, with a crossover scale $\mu$, \ie\ the physics is that of the UV fixed point at energies $E\gg\mu$, and that of the IR fixed point at $E\ll\mu$. From the long distance point of view, such a flow can be viewed as an irrelevant deformation of the IR fixed point, that becomes important at $E\sim\mu$ and is fine-tuned to approach the UV fixed point at short distances. 

Usually, RG flows of this sort are studied in flat Minkowski spacetime, $\mathbb{R}^{1,1}$, but one can ask what happens if one compactifies the spatial coordinate on a circle of radius $R$.  
The answer depends on the value of the dimensionless parameter $\mu R$. If this parameter is very large, \ie\ $\mu\gg 1/R$, the situation is as on $\mathbb{R}^{1,1}$. The energy at which the transition between the UV and IR CFT's takes place, $\mu$, is much larger than the Kaluza-Klein (KK) scale $1/R$, so the physics at energies of order $1/R$ is well described by the IR CFT. However, if this parameter is of order one or smaller, the physics at the KK scale is not well described by the IR CFT. This is the case, in particular, for the $SL(2,\mathbb{R})$ invariant vacuum of the IR CFT, which has energy $E=-c/12 R$.\footnote{If the IR CFT has states with energy $E\ll1/R$, \ie\ $L_0\simeq c_{\rm IR}/24$, they are well described by the IR CFT for $\mu\sim 1/R$. An example is Ramond ground states in a SCFT.}

The situation for the RG flow between LST in the UV and a holographic CFT in the IR described by single-trace $T\bar T$ deformed CFT, is similar to that in the last two paragraphs. The regime $\lambda\ll 1$ corresponds to $\mu R\gg 1$ in that discussion, and in this regime the effect of the deformation on the $SL(2,\mathbb{R})$ invariant vacuum of the IR CFT is negligible. This agrees with the fact that the bulk geometry \eqref{gads}, \eqref{Ngads} is well approximated by global $AdS_3$ in a wide region in $\rho$. 

On the other hand, $\lambda\sim 1$ corresponds to $\mu R\sim 1$ in the above discussion, and in that case, the effects of the irrelevant deformation on the $SL(2,\mathbb{R})$ invariant vacuum due to the single-trace $T\bar T$ deformation are not small. As we see from \eqref{zamo}, \eqref{gads}, these effects become rather dramatic for $\lambda\to 1/k$, or equivalently $R\to r_5$. This is due to the fact that the UV theory is not a CFT, like in our previous discussion, but rather a non-local theory, LST. 

It is natural to ask what happens to the bulk geometry when we approach the critical value of the coupling, $R=r_5$, and in particular, try to exceed it. Looking back at \eqref{gads}, we see that in this case the $B$-field and dilaton become complex. While the latter can possibly be cured by shifting $\Phi$ by a complex constant, the former seems problematic. To analyze the resulting background it is convenient to use a different set of coordinates, which we turn to next.

We define the coordinates $(\phi,t)$, related to $(\rho,\tau)$ in \eqref{gads} by
 \begin{equation}\label{rtorho}
     \rho=\sqrt{k}\sinh\phi~, \ \ \ \ \tau= \sqrt{\frac{k}{1+\epsilon}}t~.
 \end{equation}
 Here $\epsilon$ is related to $R$ in \eqref{gads} via
 \begin{equation}\label{alpeps}
     R=\sqrt{\frac{k(1+\epsilon)}{\epsilon}}~,
 \end{equation}
 and we set $l_s=1$ here and below. The coupling $\lambda$ in \eqref{lambda} is related to $\epsilon$ in \eqref{alpeps} via
 \begin{equation}\label{lambdaep}
     \lambda=\frac{\epsilon}{k(1+\epsilon)}~.
 \end{equation}
 Thus, $0\leq\lambda<1/k$ corresponds to $\epsilon\ge 0$; $\lambda>1/k$ corresponds to $\epsilon<-1$; and, $\lambda\leq0$ to $-1<\epsilon\leq 0$.

In the coordinates \eqref{rtorho}, the three-dimensional background \eqref{gads} takes the form
 \begin{equation}
     \begin{aligned}\label{gadscap}
         &\frac{1}{k}ds^2=-\frac{\cosh^2\phi}{1+\epsilon\cosh^2\phi}dt^2+d\phi^2+\frac{(1+\epsilon)\sinh^2\phi}{1+\epsilon\cosh^2\phi}d\varphi^2~,\\
         &\frac{1}{k}B_{t\varphi}=\frac{\sinh^2\phi}{1+\epsilon\cosh^2\phi}~,\\
         & e^{2\Phi}=\frac{kv}{p}\frac{\sqrt{1+\epsilon}}{1+\epsilon\cosh^2\phi}~.
     \end{aligned}
 \end{equation}
The background \eqref{gadscap} can be obtained\footnote{An alternative description of \eqref{gadscap} is as an $\epsilon$-dependent coset  $SL(2,\mathbb{R})\times\mathbb{R}_t/U(1)$. In this language, $\epsilon>0\; (0>\epsilon>-1)$ corresponds to axial (vector) gauging. This can be shown by generalizing the results of \cite{Giveon:1991jj,Giveon:1993ph}.}
by a $J^3\bar J^3$ deformation of the conformal sigma model on global $AdS_3$. $J^3$ and $\bar J^3$ are the holomorphic and anti-holomorphic timelike currents of the $SL(2,\mathbb{R})$ WZW model at level $k$; 
$\epsilon$ is the size of the $J^3\bar J^3$ deformation. In particular, as $\epsilon\to 0$ the background \eqref{gadscap} approaches global $AdS_3$.

For $\epsilon>0$, the background \eqref{gadscap} interpolates between global $AdS_3$ at small $\phi$ and $\mathbb{R}^{1,1}\times S^1$ with a linear dilaton in the $\phi$ direction at large $\phi$. The crossover happens at $\phi\sim\phi_*$, where $\cosh^2\phi_*= 1/\epsilon$. Thus, when $\epsilon\to\infty$, which corresponds via \eqref{lambdaep} to $\lambda\to 1/k$, the part of the geometry which looks like $AdS_3$ shrinks to zero. 

At $\lambda=1/k$, the line element in \eqref{gadscap} takes the form (after rescaling $t\to \sqrt{\epsilon}t$)
\begin{equation}\label{geo1}
\frac{1}{k}ds^2= -dt^2+d\phi^2+\tanh ^2\phi ~d\varphi^2~.
\end{equation}
The $B$-field goes to zero, and the dilaton (after an infinite shift) is given by
\begin{equation}\label{geo11}
e^{\Phi-\Phi_0}=\frac{1}{\cosh\phi}~,
\end{equation}
where $\Phi_0$ is the value of the dilaton at $\phi=0$.

The background \eqref{geo1}, \eqref{geo11} corresponds to a familiar CFT, a product of the Euclidean cigar described by the coset CFT $SL(2,\mathbb{R})/U(1)_{\rm axial}$, and 
$\mathbb{R}_t$. The level of $SL(2,\mathbb{R})$, $k$, determines the asymptotic radius of the cigar, $R=r_5=\sqrt k l_s$. We see that, as expected from the general discussion above, this background does not have any region in which it is (even approximately) described by global $AdS_3$. 

As mentioned above, $\lambda>1/k$ corresponds to $\epsilon<-1$, \eqref{lambdaep}. Looking back at eq. \eqref{gadscap}, we see that in this case, the metric on the first line describes a Euclidean space, and the $B$-field on the second line is real. The third line suggests that the dilaton is complex, but as mentioned above this can be rectified by a complex shift of $\Phi$.

The qualitative structure of the background can be seen in the original coordinates \eqref{gads} as well. As discussed above, $\lambda>1/k$ corresponds to $R<r_5$. The $B$-field on the second line is imaginary in this case; as usual in QFT and string theory, this suggests that we Wick rotate $\tau\to i\tau$, which makes the $B$-field real and the metric on the first line Euclidean. 

Coming back to the question of what happens when we increase the single-trace $T\bar T$ coupling $\lambda$ to its critical value, and try to exceed it, we arrive at the following picture. When $\lambda$ approaches $\lambda_c=1/k$, the $B$-field and $\exp(\Phi)$ go uniformly to zero, and when we try to go beyond this value, the signature of the spacetime flips, from Lorentzian to Euclidean. It is natural to interpret this as the statement that $\lambda_c$ is the maximal coupling we can turn on, and the theories that formally have $\lambda>\lambda_c$ form a decoupled branch of solutions (if they make sense at all), \eg\ in that one needs to supply a time coordinate from outside of the three-dimensional spacetime that we focus on here. 

This conclusion appears to be consistent with other known properties of the theory. In particular, as discussed in \cite{Aharony:2018bad}, $T\bar T$ deformed CFT can be formulated on a Euclidean two-torus, and one has the freedom to exchange the roles of the spatial and Euclidean time coordinates. This leads to modular invariance, which was used extensively in that paper. 

Single-trace $T\bar T$ deformed CFT also has this property, as is clear from its construction. Thus, we can interpret the spatial circle parameterized by $x$ in \eqref{ns5f1dec}, and $\varphi$ in \eqref{fullbgrho}, as Euclidean time. From that point of view, the constraint $\lambda<1/k$, or equivalently $R>r_5$, is the statement that the temperature $1/\beta$ is bounded from above by the Hagedorn temperature $T_H=1/2\pi r_5$, which is consistent with the fact that the asymptotic density of states in this theory exhibits Hagedorn growth, $S\sim \beta_H E$. Of course, the same should be true for the usual $T\bar T$ deformed CFT.

It is interesting to contrast the physics in the background $\mathcal{M}_3$ discussed in section \ref{sec2}, \eqref{fullbgrho} -- \eqref{rho0full} with $\rho_0=0$, and the deformed global $AdS_3$ \eqref{gads}, \eqref{Ngads}, discussed in this section. In the former, the region $\rho\ll R$ is described by massless BTZ, \eqref{btxmetric} -- \eqref{G3rho0} with $M=0$, for all $R$. In other words, it always approaches $AdS_3$ in the IR. On the other hand, as we discussed in this section, in the background \eqref{gads}, \eqref{Ngads} the $AdS_3$ region shrinks as we increase the single-trace $T\bar T$ coupling $\lambda$, and eventually disappears when $\lambda$ approaches its maximal value, $\lambda_c=1/k$. 
 
This is compatible with the expectation from the boundary theory, where massless BTZ corresponds to $E(0)=0$ in \eqref{zamo}, and as we see from that equation, the energy of such states does not depend on the deformation parameter $\lambda$. This agreement can of course be generalized to $M\not=0$ in \eqref{fullbgrho} -- \eqref{rho0full}. For $\rho_0>0$ in this background, the effects of the deformation are small for $\rho_0\ll R$. Using \eqref{G3rho0}, \eqref{E0}, \eqref{lambda}, this corresponds to $\lambda RE(0)/p\ll 1$. 

It is also interesting to contrast the discussion of this section with that of the system with $\lambda<0$, studied in \cite{Chakraborty:2020swe,Chakraborty:2020cgo}. The latter can be thought of as \eqref{fullbgrho}, \eqref{Nfull} with $R^2\to -R^2$. It can be generalized to a deformation of the $SL(2,\mathbb{R})$ invariant vacuum by taking $R^2\to -R^2$ in \eqref{gads}, \eqref{Ngads}. 

In the boundary theory, this leads to the appearance of complex energies when $E(0)$ exceeds a critical value, $2|\lambda|RE(0)>p$. This is reflected in the geometry as the appearance of a singularity at a large value of $\rho$, $\rho=R$. The singularity appears even if $R\gg r_5$, \ie\ the coupling $\lambda$,  \eqref{lambda} is very small. Its appearance does not disturb the $AdS_3$ region at small $\rho$, in agreement with the fact that only high energy states are influenced by its presence. 

On the other hand, the pathology discussed in this section, associated with the energy of the deformed $SL(2,\mathbb{R})$ invariant vacuum going complex, afflicts the whole spacetime \eqref{gads}, \eqref{Ngads}, as expected from the boundary point of view.

\section*{Acknowledgements} 

We thank S. Sethi for discussions. The work of SC received funding under the Framework Program for Research and
“Horizon 2020” innovation under the Marie Skłodowska-Curie grant agreement n° 945298. 
The work of AG and DK is supported in part by the BSF (grant number 2018068).
The work of AG is supported in part by the ISF (grant number 256/22). The work of DK is supported in part by DOE grant DE-SC0009924. This work was supported in part by the FACCTS Program at the University of Chicago.

 \newpage

%\bibliography{ref}\bibliographystyle{JHEP}

\providecommand{\href}[2]{#2}\begingroup\raggedright\endgroup

\end{document}